\documentstyle[prl,aps,multicol,epsf]{revtex}
\tighten
\begin{document}
\draft
\title{
Disorder and quantum fluctuations in 
superconducting films in strong magnetic fields
\\
}
\author{V. M. Galitski and A. I. Larkin}
\address{Theoretical Physics Institute, University
of Minnesota, Minneapolis, MN 55455, USA}
\maketitle
\begin{abstract}
  We find that the upper critical field in a two-dimensional
  disordered superconductor can increase essentially at low
  temperatures.  This happens due to the formation of local
  superconducting islands weakly coupled via the Josephson effect.
  The distribution of the superconducting islands is derived.  It is
  shown that the value of the critical field is determined by the
  interplay of the proximity effect and quantum phase fluctuations.
  We find that the shift of the upper critical field is connected with
  the pinning properties of a superconductor.

\end{abstract}

\begin{multicols}{2}



  The theory of the upper critical field in homogeneous
  superconductors was developed long ago \cite{Gorkov,Mak,HW}. In the
  framework of the BCS mean field theory, the value of the upper field
  is weakly temperature dependent  
  $H_{c2}(0) -  H_{c2}(T) \propto \left( T /T_{c0} \right)^2$, 
  if $T \ll T_{c0}$. 
  
  However, some experiments \cite{exp} show an anomalous upward
  curvature of the upper critical field temperature dependence as $T
  \to 0$.  A possible explanation for this behavior is based on the
  following arguments: One can not avoid having inhomogeneities in a
  superconductor. The critical field depends on disorder. Local optimal
  realizations of disorder may lead to the formation of spatial regions
  where the local upper critical field exceeds the average value.
  These regions form superconducting islands weakly coupled via the
  Josephson effect. At low temperature, proximity coupling is
  long-ranged and, thus, the global superconductivity may be
  established in the system. For the first time, this idea was
  proposed by Spivak and Zhou \cite{ZS,ZS2}. They considered
  mesoscopic effects in a dirty superconductor. The islands were
  formed due to the local fluctuations of the density of ordinary
  impurities.  It was shown that in the absence of quantum
  fluctuations, even this weak disorder leads to the macroscopic
  superconductivity at $T=0$ for an arbitrarily high external magnetic
  field.
  
  Similar phenomena at finite temperatures were considered in the work
  of Ioffe and Larkin \cite{IL}.  In this paper, the mean free path
  was supposed to be a random variable.  The optimal fluctuations
  corresponding to the appearance of superconducting islands above the
  critical temperature were found. Thermal fluctuations destroyed 
  Josephson coupling. If the energy of the Josephson interaction
  between two superconducting islands exceeded temperature, than these
  islands belonged to a same superconducting cluster. The appearance
  of an infinite superconducting cluster corresponded to the genuine
  transition point.
  
  At low temperatures, the classical thermal fluctuations are
  negligible. However, Josephson coupling can be suppressed by
  quantum fluctuations. Fluctuation-driven quantum phase
  transitions in granular superconductors were considered in 
  Refs.~\cite{Feig,SFL}.

  
  In this Letter, we study the suppression of superconductivity by the
  quantum phase fluctuations. We consider a system of superconducting
  islands in a strong perpendicular magnetic field at low
  temperatures.  We examine two different physical mechanisms of
  disorder leading to the formation of the local superconducting
  islands.  First, we study a dirty two-dimensional superconductor
  with a random distribution of impurities.  We show that in the case
  of weak disorder, mesoscopic fluctuations may become important
  only in the very vicinity (inside the Ginzburg region) of the old
  transition point, {\em i.e.}  at $H - \overline{H_{c2}}(0)\sim
  \overline{H_{c2}}(0) / g$, where $g = \sigma \,{\hbar / e^2}$ is the
  dimensionless conductance and the overline means averaging over
  disorder hereinafter. We suppose that the system possesses a large
  dimensionless conductance $g \gg 1$ and, thus, the Ginzburg region
  is very narrow.  In this small region the quantum superconducting
  fluctuations are essential, {\em i.e.} the fluctuations of the
  modulus of the order parameter must be taken into account \cite{LG}.
   
  Second, we consider a generic disordered system with a random
  diffusion coefficient:
\begin{equation}
\label{D=Dd}
 D ({\bf r}) =  \overline D  + \delta D  ({\bf r}),
\end{equation}
where a short-scale disorder characterized by the Gaussian white noise is
introduced
\begin{equation}
\label{DD}
\overline {\delta D  ({\bf r}) \delta D  ({\bf r}') } =
  \overline D ^2\, d^2 \delta\left( {\bf r} - {\bf r}'\right).
\end{equation}
The randomness can be connected with localized and extended defects
present in a superconductor. Deep into the superconducting state, this
randomness leads to the collective pinning effects. Thus, 
phenomenological constant $d$ is directly connected with the pinning
properties of a superconductor \cite{obzor} and can be extracted
independently from experiments.  Let us also note that the model can
be realized in a system of superconducting grains.  Irregularities of
grain sizes and shapes may lead to a randomness in the tunneling
constant between the granules and to a random conductance.  In this
case, constant $d$ is of the order of the grain size times a
dimensionless strength of disorder.  In the model defined by Eqs.
(\ref{D=Dd}) and (\ref{DD}), we find the shift of the upper critical field
as follows:
\begin{equation}
\label{result}
\left[ H_{c2} - \overline{ H_{c2} } \right] 
/ \overline{ H_{c2} } = {\pi \over 16}\, g \left( {d \over L_H} \right)^2,
\end{equation}
where $L_H= \left( 2 e H \right)^{-1/2}$ is the magnetic length. 

{\it Distribution of superconducting islands}. --\,
Let us consider the  vicinity of the BCS upper critical field $\overline{H_{c2}}(0)$
at zero temperature.
The transition is controlled
by the dimensionless parameter $h = \left[ H - \overline{H_{c2}}(0) \right]
/ \overline{H_{c2}}(0)$. The fluctuation region in an homogeneous superconductor is determined
by the condition $h < 1 /g$. 
We suppose that the external magnetic field is such that $h < 1$ but lies
outside the fluctuation region $h > g^{-1}$.

As we have pointed out, even above $ \overline{H_{c2}}$
superconducting islands appear. To find the distribution of the
islands, we will consider the equation for
the order parameter.  To obtain this equation, the interaction term in
the BCS Hamiltonian should be decoupled via a Hubbard-Stratonovich
field $\Delta$. Then, the one-electron degrees of freedom can be
integrated out and one gets an effective action for the
superconducting order parameter. In the vicinity of the transition, an
expansion on the order parameter is possible and we obtain the
following action:

\begin{eqnarray}
\label{Sd}
&& - S_{\Delta} = {1 \over 2} \int \Delta^*(x_1) {\cal L}^{-1}(x_1, x_2)
 \Delta(x_2) dx_1 dx_2
\nonumber \\
&&+ {1 \over 4} \int \Delta^*(x_1)  \Delta^*(x_2) 
B(\{ x_i \})
     \Delta(x_3)  \Delta(x_4) 
     \prod_i d x_i,
\end{eqnarray}
where we use $x=({\bf r},t)$ and $dx = d^2 {\bf r}\, dt$ for brevity.
In Eq.(\ref{Sd}), operator ${\hat {\cal L}}$ is the fluctuation
propagator: ${\hat {\cal L}}_\omega = \left[ \lambda^{-1} -
  \hat{\Pi}_\omega \right]^{-1}$, $\lambda$ is the BCS interaction
constant, operator $\hat\Pi$ in the coordinate representation has the
form $\Pi_\omega ({\bf r},{\bf r}') = T \sum_{\varepsilon} \Pi_\omega
({\bf r},{\bf r}';\varepsilon)$ where $\Pi_\omega ({\bf r},{\bf
  r}';\varepsilon) = {\cal G}_{\varepsilon} ({\bf r}, {\bf r}') {\cal
  G}_{\omega-\varepsilon} ({\bf r}, {\bf r}')$ and $\cal
G_{\varepsilon}$ is the Matsubara Green function.  Let us emphasize
that operator $\hat\Pi = \overline{\hat \Pi} + \delta{\hat \Pi}$
consists of a mean part and a random part $\delta{\hat\Pi}$ which is
responsible for the effects under consideration.  Non-linear operator
$B$ in Eq.(\ref{Sd}) corresponds to the  diagrams calculated
explicitly by Maki \cite{Maki} and Caroli {\em et al.} \cite{deG}.  In
the vicinity of the transition we can neglect the randomness in the
$\Delta^4$ term.

The saddle point approximation
${\delta S / \delta \Delta({\bf r},t)} = 0$ results in the
time-dependent Ginzburg-Landau equation for a gapless superconductor
\cite{GorE}.  When considering the spatial distribution of the islands
we can disregard dynamic effects and consider the static form
of the Ginzburg-Landau equation.

If we neglect randomness in the Cooperon, indeed, there are no
non-trivial solutions for the corresponding mean-field equation above
the BCS upper critical field $h > 0$.  However, if the random part of
kernel $\delta{\hat \Pi}$ possesses an eigenvalue which is greater
than $h$ a non-trivial solution appears.  This corresponds to the
appearance of a local superconducting island.

To find the distribution of the islands, one should find the
distribution function of the eigenvalues for the random operator
${\hat \Pi}$:
\begin{equation}
\label{dC}
{1 \over \nu} \int\Pi({\bf r}, {\bf r}') \,\psi({\bf r}') d^2{\bf r}' = 
\left( \overline{\epsilon}  + \delta\epsilon \right) \psi({\bf r}),
\end{equation}
where $\epsilon= \overline{\epsilon} + \delta\epsilon$ is the
dimensionless eigenvalue of the Cooperon and $\nu$ is the density of
states per spin at the Fermi line. In the absence of a randomness in
$\hat \Pi$, the eigenvalues $\overline{\epsilon}$ are well-known.  The
spectrum is discrete and is parameterized by the
Landau level indexes.  The random part smears out the eigenvalues.

The ``density of states" can be defined as
\begin{equation}
\label{dos}
\rho(\epsilon) = \int  {\cal D}  \left\{ \delta\Pi \right\}
\delta \left( \epsilon - \epsilon\left[ \Pi \right] \right) w\left[ \delta\Pi \right],
\end{equation}
where $w\left[ \delta\Pi \right]$ is the distribution function for
the Cooperon which is supposed to be Gaussian with correlator
$\overline{ \delta\Pi({\bf r_1},{\bf r_2}) \delta\Pi^*({\bf r_3},{\bf r_4}) }$.

To find the density of states $\rho(\epsilon)$, we use the optimal
fluctuation method \cite{Nar}. This means that we evaluate
functional integral (\ref{dos}) in the saddle-point approximation. Let
us note that the problem of finding $\rho(\epsilon)$ is analogous to
the problem of density of states of a particle in a random potential.
In the presence of an external magnetic field in 2D, the problem is
simplified, since the coordinate dependence of the wave functions is
dictated by the magnetic field \cite{IL}.

In the case $\overline{ \left( \delta\epsilon \right)^2 } \ll \left(
  \delta\epsilon \right)^2 \ll 1$, the solution has a form of rare
islands. In the vicinity of a spherically symmetric island located at
a point ${\bf r}_i$, the ``wave function'' can be taken in the
following form:
\begin{equation}
\label{psi}
\psi_i ({\bf r}) = {1 \over \sqrt{2 \pi} L_H}\, \exp{
\left\{ -{ \left( {\bf r} - {\bf r}_i \right)^2 \over 4 L_H^2} \right\}
}.
\end{equation}
In the first approximation we obtain:
$$
\delta\epsilon = {1 \over \nu} \delta\Pi_{00} \equiv
{1 \over \nu} \int \psi_i({\bf r_1}) \delta\Pi({\bf r}_1,{\bf r}_2) \psi_i({\bf r_2})
d^2{\bf r}_1 d^2{\bf r}_2.
$$
The distribution function reads:
\begin{equation}
\label{ro}
\rho(\epsilon) \propto \exp{ \left[- { \left(\delta\epsilon\right)^2 \over 2  I}
\right]},
\end{equation}
where $I = \overline{ \delta\Pi_{00}^2}/ \nu^2$. 

In the case of a dirty metal, correlator $I$ can be calculated with
the help of the conventional cross diagram technique \cite{ZS}. This
yields the following estimate for the correlator: $I_1^{-1} \sim g^2$,
where $g$ is the dimensionless conductance and index ``1'' refers to
the first model we consider (weak mesoscopic fluctuations in a dirty
metal).

In the system with a short-scale randomness in the diffusion
coefficient (\ref{DD}), we can calculate the correlator using the
differential equation for the Cooperon which in the presence of an
external magnetic filed has the form:
\begin{equation}
\label{eqcop}
\bigl[ \bbox{\partial} \,  D ({\bf r})\,
\bbox{\partial}  + i \varepsilon \bigr]
\Pi ({\bf r}, {\bf r'};\varepsilon) 
 = \left(2 \pi \nu\right)\, \delta({\bf r} - {\bf r'}),
\end{equation}
where $\bbox{\partial} = -i \bbox{\nabla} - 2 e {\bf A}({\bf r})$.
One can solve Eq.(\ref{eqcop}) using a simple perturbation theory with
respect to $\delta D $.  With the help of Eq.(\ref{DD}), we get the
correlator and find the distribution function which can be written in
the form (\ref{ro}) with $I_2^{-1} = 8 \pi \left( {\displaystyle L_H
    / \displaystyle d } \right)^2$.

The modulus of the order parameter in a superconducting island is
random and parameterized by random variable $\epsilon$ (see
Eqs.(\ref{dC}) and (\ref{ro})).  Using the explicit expression (see
Ref.~\cite{deG}) for the non-linear operator $B$ in Eq.(\ref{Sd}), one
can get the following ``mean-field'' value of the order parameter for
a spherically symmetric island:
\begin{equation}
\label{modD}
\left| \Delta_0 \right| 
=
\sqrt{4 \pi} \, {\overline D  \over L_H} 
\sqrt{\delta\epsilon - h}
\end{equation}
and the coordinate dependence of the order parameter is described by
$\Delta_i ({\bf r}) = \Delta_0 \psi_i ({\bf r})$, where $\psi_i({\bf r})$ is
defined in Eq.~(\ref{psi}). Let us note that the typical size of a
superconducting island is $L_H$. The typical distance between the
islands is exponentially large $R \sim L_H \exp{\left[ h^2 / 4 I
  \right] }$.

{\it Josephson coupling}. -- For each realization of disorder, there
is a fixed spatial distribution of the superconducting islands. The
interaction Hamiltonian for such a system can be obtained from
Eq.(\ref{Sd}) and has the standard form
\begin{equation}
\label{HJ}
{\cal H}_{\em int} = \sum\limits_{ij} J_{ij} \cos{
\left( \phi_i - \phi_j + A_{ij} \right) },
\end{equation}
where $J_{ij}$ is the Josephson energy of the interaction between
islands $i$ and $j$ and $A_{ij}$ is the phase-shift due to the
magnetic field.  The average value of the Josephson energy is
$\overline{J(R)} \propto R^{-2} \exp{\left[-R/L_H \right]}$.  Since
the typical distance between the islands is exponentially large
compared to $L_H$, the average Josephson energy is negligible.
However, as it was shown in \cite{ZS}, the variation of the Josephson
energy decays as a power law only $\overline{J^2(R)} \propto R^{-4}$

The Hamiltonian (\ref{HJ}) describes a ``frustrated'' two-dimensional
$XY$-model with random bonds. The frustration comes both from the
Josephson energy which is random and from the phase difference due to
the magnetic field. At zero temperature such a system should show a
glassy behavior if there are no effects capable of destroying phase
coherence between the islands.

{\it Transition point}. --\, To find the transition point, we use
action (\ref{Sd}) which describes the dynamics of the superconducting
order parameter.

Let us present the superconducting order parameter in the following
form
\begin{equation}
\label{Ds}
\Delta({\bf r},t) = \sum_i\, \left| \Delta_{0\, i} \right|
\psi_i \left( {\bf r} \right)
e^{i \phi_i(t)},
\end{equation}
where $\psi_i$ is defined in (\ref{psi}). We consider the islands in
which the modulus of the order parameter is fixed by the static
mean-field equations (\ref{modD}) and only the phase is allowed to
fluctuate. Finally, we obtain the following action describing the
system of local superconducting islands:
\begin{eqnarray}
\label{S}
S = -\int dt \sum_{i} \nonumber 
 \Biggl\{ \Biggr. \int dt'
\eta_i  {
\cos{ \left[ \phi_i(t) - \phi_i(t') \right] }
\over \left( t - t' \right)^2 
} -
{1 \over E_c}
\left( { \partial \phi_i \over \partial t } \right)^2  \nonumber \\
 + \sum_{j \ne i}
 \int dt' 
  J_{ij}(t-t') \cos{ \left[ \phi_i(t) - \phi_j(t') + A_{ij} \right] }
\Biggl. \Biggr\}.
\end{eqnarray}
The coefficient in the dissipative term is random and connected with
the modulus of the order parameter (\ref{modD}) in an island $\eta_i =
{ \nu \left| \Delta_{0i} \right|^2 / 8 \pi e \overline D H}$.  Note
that the typical value of the coefficient is large $\eta \sim g h$.

We keep the $\omega^2$ term in the action.  As we will see below, the
effective charging energy appears only as a high-frequency cut-off.
With the logarithmic accuracy, the exact value of $E_c$ is not
important in our problem.

Let us integrate out the high-frequency degrees of freedom in the
action.  First, we consider strong enough magnetic fields so that the
network of the superconducting islands is very dilute and the average
Josephson energy is exponentially small. In the domain $\omega \gg J$,
only the first two terms in action (\ref{S}) are important.  In this
case, the action can be written as a sum of single-island actions.  The
phases in different islands fluctuate independently.  With the aid of
the single-island action one can integrate out the high-frequency phase
fluctuations using the renormalization group developed by Kosterlitz
for a spin system with long-range interactions \cite{Koster}.  Since the
first term in Eq.(\ref{S}) is not Gaussian, coefficient $\eta$ gets
renormalized when integrating out fast variables.  The corresponding
renormalization group equations are identical to the ones derived in
\cite{Feig,Koster}. The solution of the RG equation for the renormalized
``viscosity'' coefficient has the following simple form:
\begin{equation}
\label{RG}
\eta(\omega) = \eta - {1 \over 2 \pi^2} \ln{E_c \over \omega}.
\end{equation}
Note that this equation is valid unless $\eta(\omega)$ becomes of the
order of unity. This happens at times $t_c \sim \omega_c^{-1} \sim
E_c^{-1} \exp{\left[ 2 \pi^2 \eta \right]}$.  At larger times the
phase fluctuates rapidly.

If the Josephson interaction between two islands is such that $J_{ij}
t_c^{\rm min} < 1$, then the Josephson term does not affect the phase
dynamics at any times and can be treated as a small perturbation.
However, there is always a finite probability of finding a pair of
islands for which $J_{ij} t_c^{\rm min} > 1$. In this case, $J_{ij}$
stabilizes the fluctuations of the relative phase $\left[ \phi_i(t) -
\phi_j(t) \right]$ and the corresponding critical time increases $t_c \sim E_c^{-1}
\exp\left[ 2 \pi^2 \left( \eta_i + \eta_j \right) \right]$.

Recall, that $\overline{J_{ij}^2} \propto R_{ij}^{-4}$ and the
probability distribution for random quantity $t_c$ is known
and determined by Eqs.~(\ref{ro}) and (\ref{modD}).
Thus, one finds the probability of finding a  pair of strongly 
correlated superconducting islands:
\begin{equation}
\label{P}
P \sim \exp{ \left[ -2 \pi^2 g \left( h - {\pi^2 \over 2} g I \right) \right]}.
\end{equation}
At large fields, this probability is exponentially small.  As the
external magnetic field decreases, the fraction of the strongly
correlated superconducting islands increases and finite size
superconducting clusters are formed. At some threshold field, the
infinite superconducting cluster appears and it corresponds to the
macroscopic superconductivity.  We can estimate the location of the
transition point as a field at which probability (\ref{P}) is of
the order of unity.  This yields
\begin{equation}
\label{hc2}
h_{c2} = {\pi^2 \over 2}g I. 
\end{equation}

Let us mention that result (\ref{hc2}) can be obtained more formally
by calculating correlator $C = \int_0^{\infty} \left\langle
  \exp{\left[ i \phi_j(t) - i \phi_j(0) \right]} \right\rangle_S dt$.
At the transition point, the correlator diverges. One can perform the
virial expansion with respect to the density of islands in $C$. As in
the theory of liquids and gases and in the theory of spin-glasses with
RKKY interactions \cite{Meln}, the virial expansion can not prove the
very existence of the transition.  However, it determines the
transition point if there is one.  The transition is defined as a point
at which all terms of the virial expansion become of the same order.
Comparing the contribution in correlator $C$ from independent islands
and the one from pairs, we find Eq.~(\ref{hc2}).

In the case of the weak mesoscopic disorder $I_1 \sim g^{-2}$ and the
shift of the upper critical field, if any, is small: $h_{c2} \sim
g^{-1}$.  Let us note that this result may acquire some logarithmic
corrections of the order of $\left( g^{-1} \ln{g} \right)$ which are,
however, beyond the scope of our investigation. Within the logarithmic
accuracy, we can not distinguish the mesoscopic effects under
consideration from the usual superconducting fluctuations inside the
Ginzburg fluctuation region.

In the case of strong disorder, the shift of the critical region can be
large and it is described by Eq.(\ref{result}).  Phenomenological
constant $d$ measures the strength of disorder which can be connected
with dislocation clusters, grain boundaries in polycrystal samples
etc.  Let us emphasize, that the pinning parameters in a
superconductor are determined by $d$. For example, the critical
current of a superconducting film \cite{Ovc} in the collective pinning regime is
$j_c/j_{c0} \approx \left[ {H_{c2}(0) \over H
    }\right]\, d^2/L^2_{H_{c2}}$, where $j_{c0}$ is the depairing current
in zero field.  Let us also note that a possible randomness in the BCS
interaction constant $\lambda$ would lead to the same effects on the
upper critical field.

At finite temperatures, there is no phase transition in a strict
sense. At any fields, a finite, though exponentially small, resistance
exists in a two-dimensional superconductor if a magnetic field
is applied.  One can define the upper critical field as a field at
which a sharp fall in the resistance takes place.  At very low
temperatures, $h_{c2}$ is determined by Eq.~(\ref{result}).  As the
temperature increases, $h_{c2}(T)$ decreases very rapidly.  First of
all, at a finite temperature, the Josephson coupling decays
exponentially at the distances larger than $\sqrt{D / T}$. Second, the
thermal fluctuations destroy the Josephson coupling at $J_{ij} \sim
T$.  The both effects lead the following estimate of the transition
temperature $T_c(h) \sim T_{c0} \exp{\left[ -h^2/4 I \right]}$. Thus, in a
relatively wide region $\sqrt{I} < h < h_{c2}(0)$, the critical
temperature depends on the external magnetic field exponentially.

The increase of $H_{c2}$ at low temperatures has been observed in a
number of experiments. The mechanism proposed in the
present paper can give a possible explanation for the effect. To
reveal whether or not the fluctuation effects in disorder are
responsible for the increase of $H_{c2}$, it would be interesting to
investigate possible correlations between the value of the upper
critical field at low temperatures and the pinning properties.

This work was supported by NSF Grant DMR-9812340. The authors are
grateful to L. Glazman and B. Spivak for valuable discussions.

\end{multicols}
\end{document}